\documentclass{article}
\begin{document}
\baselineskip18pt

\title{Dynamics of Uncertainty in  Nonequilibrium  Random Motion}
\author{Piotr Garbaczewski\\
Institute of Physics,  University  of Zielona  G\'{o}ra, 65-516
Zielona G\'{o}ra, Poland\thanks{Presented at the XVII Marian
Smoluchowski Symposium on Statistical Physics, Zakopane, Poland,
September 4-9, 2004}} \maketitle
\begin{abstract}
Shannon  information  entropy  is a natural measure of  probability
 (de)localization and thus (un)predictability in  various procedures
  of data  analysis for  model systems. We pay  particular attention to
links between the Shannon   entropy and  the related
Fisher information  notion,  which  jointly  account for  the shape
and extension of   continuous probability distributions.
Classical, dynamical and  random  systems  in general  give rise to
time-dependent probability  densities  and  associated information
measures. The induced dynamics of Shannon and Fisher  functionals
 reveals   an  interplay among various characteristics
 of  the  considered  diffusion-type systems: information, uncertainty and
 localization while put  against  mean energy and its  balance.
\end{abstract}

\section{Introduction}

We shall investigate  relationships   between   the  dynamical features
of the  differential entropy  (Shannon  entropy  of general time-dependent continuous
probability densities),  \cite{shannon,sobczyk},  and so-called   hydrodynamical
conservation laws (mass/probability, momentum and energy balance
in the mean) of the corresponding (ir)reversible diffusion-type process.

In part, our arguments  derive from a  standard
trajectory interpretation in  which  random transport is modelled
in terms of  a Markovian  process  and  its  sample paths.
The pertinent  process  obviously  complies with the  Fokker-Planck dynamics of an initially  prescribed
 probability density, \cite{lasota,risken}.

However, we would like  to point out that   a generic property of  physically
interesting  cases is  their  conflict with rather stringent
  growth and  H\"{o}lder  continuity   restrictions   for
 drift and diffusion coefficient functions.  Those bounds need to be respected  for a
  mathematically consistent  definition   of the   process  and  its  transition density functions.
 In most of "typical" cases, the   uniqueness   and non-explosiveness of the process   cannot
   be guaranteed, see however \cite{qian,eberle} how to evade the explosiveness problem.

This formal defect of a theoretical framework is usually
 bypassed in  a pragmatic computer-assisted  research  by  neglecting
the unwanted (even if annoying, interpreted as artifacts)   contributions
to  the data.  In view of  low probability for  troublesome
events (explosive behavior), it is often taken for
granted that a mathematical pedantry is  here  unnecessary   and
that  the Langevin equation  can be employed in the study of
diffusion-type processes  without any specific precautions. This attitude
is omnipresent,    when one attempts  to solve   explicitly the  "obvious" Fokker-Planck or
Smoluchowski diffusion equation, but  does \it not \rm  inquire into an issue of
transition probability density functions.
Needless to say, with the latter step ignored,  the random variable  and random  path notions are
often  maintained as  legitimate elements of the analysis.

On the other hand,  the previously
mentioned  restrictions on drift and diffusion coefficient
functions  may   be relaxed  in a controlled way  to allow for a
consistent theory. There is an  obvious price  to be paid,  one
should  admit and learn to live with  non-unique  and possibly explosive
stochastic processes, all of them being capable to drive accordingly a
\it unique \rm   probability density.
Examples of  such   milder (than usual) restrictions can be found in
Refs. \cite{qian,qian1,eberle}.

The previous obstacles motivate a   principal peculiarity of our approach
which  is rooted in the fact that we  extract  relevant data exclusively
 from   the  (basically, spatial) probability density of the pertinent
 dynamical process and this density gradient, with no explicit mention of
 random or deterministic paths.  Clearly,  there   are many distinct stochastic
  processes which can be associated with the once  prescribed   Fokker-Planck
  dynamics of a concrete  probability density.

It is widely accepted in the literature  to invoke relative
Kullback-Leibler entropies  as "distance measures" in the set
of   different   probability densities. In particular,   for  comparison
of different solutions of a given  Fokker-Planck equation, \cite{kullback,risken}.
One often takes for granted that  the Kullback entropy is a proper
  analog of Boltzmann's H-function in the diffusion process setting.
  The reason is that it never takes negative values,  while
the differential  entropy does. Its time rate is negative, hence
refers to  a continually decreasing function in accordance with thermodynamical intuitions,
 which is not necessarily the case for  Shannon entropies, \cite{shannon}.

There is one minor obstacle: a  closer inspection  shows that the
Kullback entropy is  mainly explored under standard severe
  restrictions upon drift and diffusion coefficients.
See e.g. \cite{risken} for a verbal statement: "we assume that the
drift coefficients   have no singularities   and that they do not
allow the solutions to run away to infinity".  Not surprisingly,
in a statistical physics lore, a tacit assumption  is  that "all
solutions of the Fokker-Planck equation finally agree if we wait
long enough". Hence it is believed to be  immaterial to discuss
their behavior in other regimes,  than close-to-equilibrium
(asymptotic  invariant density).

Another peculiarity of our approach is that
we are not quite interested in  "measuring a distance"  between  two  different
  probability densities. We  rather wish to make a  comparison of   the very \it same \rm
 non-equilibrium density and its  differential entropy    at different
 stages of their  time evolution. In particular, the  difference  of the  respective
  entropy values  at two time instants is a legitimate  "distance measure" (information
  gain or  loss), \cite{shannon}, the   time rate of information entropy is also  a  well
  defined quantity.
For those reasons, we   deliberately avoid the use of the Kullback
entropy and insist on  investigating  the role and potential utility
 of the Shannon-type information entropy per se.

 Other motivations come from varied attempts to use information
 theory concepts as natural tools for quantifying signatures of disorder
 and  its intrinsic dynamics (time  rate of generation/propagation of
disorder, information flow, entropy production rate). This
involves  an issue of non-equilibrium steady states  and  the time
rate ("speed") of  an asymptotic approach to equilibrium, when
time-reversible stationary processes   ultimately enter the game,
\cite{dynamics}.
Discussions \cite{gaspard,gaspard1,rondoni} of a
physical role of the probability density gradient in classical
non-equilibrium thermodynamics of irreversible processes are  worth
mentioning, to place our discussion in  a proper context.

An analysis of  links \cite{nicolis,mackey} between dynamical
systems, weak noise and information entropy production is also
useful to that  end. An independent  input comes from general
studies  of the  dynamical origin of increasing entropy ("dynamical
foundations of the evolution of entropy to maximal states"),
entirely  carried out with respect to time-dependent probability
densities, \cite{lasota,mackey,nelson} see also \cite{huang,risken}.

The  intertwined  dynamics of the differential (information)  entropy and  t
he probability localization  properties  (dynamics of uncertainty)  appears to be
 an intrinsic  physical feature of  any  formalism operating with general
  time-dependent (in the present paper, spatial) probability distributions.

\section{Information entropy  and its dynamics}

Let us  consider a classical dynamical system in $R^n$  whose evolution
is governed by equations of motion:
\begin{equation}
\dot{x} = f(x) \label{set}
\end{equation}
where $\dot{x}$ stands for the time derivative and $f$  is an  $R^n$-valued
function of $x\in R^n$,  $x= \{ x_1,x_2,...,x_n\}$.
The statistical ensemble of solutions  of such dynamical equations   can be
 described by a time-dependent probability density  $\rho (x,t)$ whose dynamics
 is given by the generalized  Liouville (in fact, continuity) equation
 \begin{equation}
\partial _t \rho = - \nabla \cdot (f\, \rho )
\end{equation}
where $\nabla \doteq \{ \partial /\partial x_1,...,\partial /\partial x_n\} $.

With any continuous probability density $\rho \doteq \rho(x,t) $,  where $x\in
R^n$ and  we allow for  an explicit time-dependence, we can
associate a probability density functional  named Shannon entropy of a continuous probability
distribution (convergence of an integral is presumed), \cite{shannon}:
\begin{equation}
{\cal{S}}(\rho ) =  - \int \rho \, \ln \rho \,  dx \, .
\end{equation}

In general,  ${\cal{S}}(\rho ) \doteq  {\cal{S}}(t)$ depends on time.  Let us  take for granted that
an interchange of time derivative  with  an indefinite  integral is allowed (suitable
precautions are necessary with respect to  the convergence of integrals). Then, we readily get
an identity, \cite{andrey,plastino,nicolis}:
\begin{equation}
\dot{\cal{S}} =  \int \rho \, (div \, f) dx \doteq \langle \nabla \cdot f \rangle \label{Liouville}\, .
\end{equation}
Accordingly, the information entropy ${\cal{S}}(t)$ grows with time only if the dynamical
system has  positive  mean   flow divergence.

However, in general  $\dot{\cal{S}}$ is  not positive definite.
For example, dissipative dynamical  systems  are characterized by the   negative
(mean) flow divergence.
Fairly often, the divergence of the flow  is constant, \cite{andrey}. Then,  an "amount of
information" carried by a  corresponding  statistical ensemble (e.g. its density)  increases,  which is
 paralleled by  the information  entropy decay (decrease).

An example of a system with a point  attractor  (sink) at origin is a one-dimensional
 non-Hamiltonian  system $\dot{x}= - x$.  In this case  $div f = -1$ and $\dot{\cal{S}} = -1$.
 A discussion  of dynamical systems  with strange (multifractal) attractors, for which
 the Shannon information entropy  decreases indefinitely (the pertinent steady states are
 no longer represented by probability density functions) can be found in \cite{andrey,nicolis}.

 We note that for Hamiltonian systems,
  the phase-space flow is divergenceless, hence $\dot{\cal{S}} =0$ which implies that "information
   is conserved" in Hamiltonian dynamics.
Take for example a two-dimensional conservative system with $\dot{x}=p/m$ and
$\dot{p}= (-\nabla V)$, where  $H=p^2/2m + V(x)$.  The classical equations of motion
yield the standard  Liouville equation  (which is a special case of  Eq.~(\ref{Liouville})):
\begin{equation}
{\frac{\partial}{\partial t}} \rho = - {\frac{p}{m}} {\frac{\partial }{\partial x}}  \rho + (\nabla V)
{\frac{\partial }{\partial p}}  \rho
\end{equation}
 for the phase-space density $\rho (x,p)$. The corresponding divergence  vanishes and the
 phase space volume is conserved.
For non-Hamiltonian systems we may generically  expect the phase-space volume contraction,  expansion or both
at  different stages of time evolution,
\cite{andrey,nicolis}.

In case of a general  dissipative dynamical system (\ref{set}), a controlled  admixture
of noise  can stabilize dynamics and yield asymptotic invariant densities.
For example, an additive modification of
the right-hand-side of   Eq.~(\ref{set}) by white noise term $A(t)$ where
$\langle A_i(s)\rangle =0$ and $\langle A_i(s)A_j(s')\rangle = 2q \delta (s-s') \delta _{ij}$,
 $i=1,2,...n$, implies the Fokker-Planck-Kramers equation:
 \begin{equation}
\partial _t \rho = - \nabla \cdot (f\, \rho )  + q \Delta \rho
\end{equation}
where $\Delta \doteq \nabla ^2 = \sum_{i} \partial ^2  /\partial x_i^2$.
 Accordingly, the differential entropy dynamics  would take another form than this  defined by
  Eq.~(\ref{Liouville}):
\begin{equation}
\dot{\cal{S}} =  \int \rho \, (div \, f) dx   + q \int {\frac{1}{\rho }}  \label{daems}
 (\nabla \rho )^2\, dx   .
\end{equation}
Now, the     dissipative  term  $\langle \nabla \cdot f \rangle <0 $ can be counterbalanced
by a strictly positive    stabilizing contribution  $q \sum_{i}  \int {\frac{1}{\rho }}
 (\partial \rho /\partial x_i)^2\, dx$. This allows
  to expect that, under suitable circumstances  dissipative systems with noise may
  yield  $ \dot{\cal{S}} = 0$.  In case of   $\langle \nabla \cdot f \rangle \geq 0 $,
   the information  entropy would    grow monotonically.

At this point,  we  depart from an  explicit phase-space background  for further
 discussion and   consider  exclusively  \it  spatial  \rm  Markov diffusion
processes with a diffusion coefficient $D$  (constant or time-dependent, with  standard
dimensions of $k_BT/m\beta $ where $\beta $ is a friction coefficient, or ${\hbar }/2m$).
We  admit them to drive space-time inhomogeneous probability densities $\rho
= \rho (\overrightarrow{x},t)$ with $\overrightarrow{x}\in R^3$.
The density gradient is introduced
in conjunction with so-called osmotic velocity field
$\overrightarrow{u} = D \overrightarrow{\nabla } \ln \rho $, c.f.
\cite{nelson}. The probability density is to obey  the continuity equation, with
 $\overrightarrow{v}$
set in correspondence with the previous vector-valued function $f \in R^n$:
\begin{equation}
\partial _t \rho = - \overrightarrow{\nabla } \cdot (\overrightarrow{v}\, \rho )
\end{equation}
where a (postulated) decomposition:  $\overrightarrow{v}(\overrightarrow{x},t) =
\overrightarrow{v} \doteq
\overrightarrow{b} - \overrightarrow{u}$ allows us  to infer  the  related Fokker-Planck equation:
\begin{equation}
\partial _t \rho  = D \Delta \rho - \overrightarrow{\nabla }\cdot (\overrightarrow{b} \rho )
\end{equation}
with a forward drift  function  $\overrightarrow{b}(\overrightarrow{x},t)$.

To make things simpler, we assume to have given   a concrete  functional expression
 for the  time-independent   forward drift $\overrightarrow{b}(\overrightarrow{x})$
 (here, we do not bother about its detailed justification  on phenomenological  or model
 construction grounds)  and fix
 initial/boundary data for  the probability density $\rho $.
We shall not demand the validity of   standard mathematical restrictions
(growth and H\"{o}lder continuity conditions), guaranteeing the existence of
non-explosive solutions $\overrightarrow{X}(t)$ of the underlying  stochastic differential equation,
since that would exclude a vast number of physically interesting situations, when the
corresponding partial differential (Fokker-Planck)  equation nonetheless \it  has \rm well
 defined solutions of the initial/boundary value problem.
Therefore we prefer to investigate random diffusive motion in terms of probability
densities, and \it not \rm directly in terms of  paths  (sample trajectories) induced by
  random variable  $\overrightarrow{X}(t)$.

 With a solution  $\rho (\overrightarrow{x},t)$ of  the Fokker-Planck equation, we
  associate its  differential (Shannon   information)   entropy \rm  ${\cal{S}}(t)= - \int \rho \,
   \ln \rho \,  d^3x $ which typically is not time-independent, \cite{nicolis,andrey}.
The  evolution (dynamics of information)  and  rate of change in  time  of the
entropy ${\cal{S}}$  directly   follow.

First, let us notice that in the particular  case of $\overrightarrow{v}=-
\overrightarrow{u}$  (i.e. $\overrightarrow{b}=0$), where $\overrightarrow{u} =
D \overrightarrow{\nabla } \ln \rho $,   we infer the standard free Brownian
motion outcome, \cite{gaspard}:
\begin{equation}
{\frac{d{\cal{S}}}{dt}} = D\cdot \int
{\frac{(\overrightarrow{\nabla } \rho )^2}{\rho }} d^3x > 0
\end{equation}
to be compared with the previously introduced  stabilizing term in Eq.~(\ref{daems}).
Thus, information   entropy  definitely   increases in the Brownian
motion and its time rate may be interpreted as  the rate of
information decay  (uncertainty increase) in the course of the diffusion process, in  close parallel  with the
casual perception of the laws of thermodynamics.

While passing  from the free Brownian motion to the forced one and more general diffusion-type processes,
we  shall demand  the current velocity
$\overrightarrow{v}(\overrightarrow{x},t)$  to be  a
 gradient field $\overrightarrow{v} \doteq  \overrightarrow{b} -
\overrightarrow{u}$,  where the  forward drift
$\overrightarrow{b}(\overrightarrow{x},t)$  of the process   may be time-dependent.

Boundary restrictions  upon  $\rho $, $\overrightarrow{v} \rho $ and
$\overrightarrow{b} \rho $ to  vanish at spatial infinities (or  at
finite spatial volume boundaries) yield the information  entropy balance equation:
\begin{equation}
\frac{d{\cal{S}}}{dt}  = \int [\rho \, (\overrightarrow{\nabla
}\cdot \overrightarrow{b})
 + D \cdot  {\frac{(\overrightarrow{\nabla }\, \rho )^2}\rho }]\, d^3x
\end{equation}
to be compared with the previous, vanishing  $\overrightarrow{b}$,
case.  We can  rewrite  this equation  as follows:
\begin{equation}
D \dot{\cal{S}} \doteq  \langle  \overrightarrow{u}^2\rangle +
D \langle \overrightarrow{\nabla }\cdot \overrightarrow{b} \rangle    =
D\langle  \overrightarrow{\nabla }\cdot
 \overrightarrow{v}
 \rangle
\end{equation}
or equivalently
\begin{equation}
D \dot{\cal{S}} = \langle \overrightarrow{v}^2\rangle  - \langle \overrightarrow{b}\cdot
\overrightarrow{v} \rangle =
- \langle \overrightarrow{v}\cdot \overrightarrow{u} \rangle \, . \label{major}
 \end{equation}
 Note  that  we have employed an identity
 \begin{equation}
 \langle \overrightarrow{u}^2\rangle =
 - D \langle \overrightarrow{\nabla } \cdot \overrightarrow{u} \rangle \, .
 \end{equation}
 The  osmotic velocity field, by its very  definition, always  has
  negative mean divergence.

The mean divergence of the current velocity field has no definite sign.
Therefore the    monotonic increase
 of ${\cal{S}}(t)$ is guaranteed only if
$\langle  \overrightarrow{\nabla }\cdot
 \overrightarrow{v} \rangle > 0$, or equivalently $\langle  \overrightarrow{\nabla }\cdot
 \overrightarrow{b} \rangle > \langle  \overrightarrow{\nabla }\cdot
 \overrightarrow{u} \rangle  $.
Invariant probability densities   are allowed    when the information entropy  remains constant in
time: $\frac{d{\cal{S}}}{dt}=0$, that is when  $\langle  \overrightarrow{\nabla }\cdot
 \overrightarrow{v} \rangle = 0$, i. e. $
 \langle  \overrightarrow{\nabla }\cdot
\overrightarrow{b} \rangle  =  \langle \overrightarrow{\nabla }\cdot
 \overrightarrow{u} \rangle $.

The simplest    realization of the state of  equilibrium   is granted by
 $\overrightarrow{b}=\overrightarrow{u} = D \overrightarrow{\nabla } \ln \rho
 $,  when the diffusion current identically
vanishes:  $\overrightarrow{v}=\overrightarrow{0}$.
For familiar  Smoluchowski diffusion processes whose  drifts have the form
$\overrightarrow{b}= - (1/m\beta ) \overrightarrow{\nabla } V$,
 where $V$ is time-independent, we immediately arrive at the
classic  equilibrium identity
\begin{equation}
-(1/k_BT)\overrightarrow{\nabla }V =
\overrightarrow{\nabla }\ln\, \rho
\end{equation}
 with the  implicit Einstein fluctuation-dissipation formula
$D=k_BT/m\beta $ ($k_B$ is the Boltzmann constant).

It is not  obvious at all that  the differential  (Shannon information)  entropy needs to
 increase,  when  a given "attracting"    state of equilibrium (invariant density) is being
 asymptotically  approached.  Entropy decay scenario seems to be  equally likely in this situation.

A hint to  this end:  invoking the standard Smoluchowski diffusion,  fix $\overrightarrow{b}(\overrightarrow{x})$,
 i. e. external force, and fine-tune an  initial density
   $\rho_0(\overrightarrow{x})$ so that $\langle  \overrightarrow{\nabla }\cdot
 \overrightarrow{b} \rangle < \langle  \overrightarrow{\nabla }\cdot
 \overrightarrow{u} \rangle  $ and therefore $\langle  \overrightarrow{\nabla }\cdot
 \overrightarrow{v} \rangle < 0$. Realization: consider the    one-dimensional  example
  with $b(x)= -\gamma x $, $\gamma >0$ and  choose
  $\rho _0(x)= [1/(\sigma \sqrt{2\pi }] \exp[- x^2/2\sigma ^2]$,  implying
$\langle \nabla \cdot u \rangle = - D/\sigma ^2$. Finally adjust $\sigma $  and/or $\gamma $
to yield $D/\sigma ^2 < \gamma $.

Let us also observe that, in view of $D \dot{\cal{S}}=-
\langle \overrightarrow{v}\cdot \overrightarrow{u} \rangle $,
by reintroducing the  diffusion current $\rho \, \overrightarrow{v}$ and recalling that
 $\overrightarrow{u} = (D  \overrightarrow{\nabla }\rho )/\rho $, we arrive at:
\begin{equation}
D \frac{d{\cal{S}}}{dt} =  - \int [\rho ^{-1/2}(\rho
\overrightarrow{v})]\cdot [ \rho ^{-1/2}(D \overrightarrow{\nabla
}\rho )] d^3x
 \, .
 \end{equation}
By means of the Schwarz inequality we  infer  an upper bound
on the magnitude of the  information  entropy time rate:
\begin{equation}
D |\frac{d{\cal{S}}}{dt}| \leq \left<
\overrightarrow{v}^2\right>^{1/2} \left<
\overrightarrow{u}^2\right>^{1/2}\, .
\end{equation}
As a byproduct we realize  that  a necessary
condition for $\frac{d{\cal{S}}}{dt}\neq 0$ is that  \it both \rm
$\left< \overrightarrow{v}^2\right>$  and $\left<
\overrightarrow{u}^2\right>$ are nonvanishing.
A sufficient condition
for $\frac{d{\cal{S}}}{dt}= 0$ is that \it
any \rm of $\left< \overrightarrow{v}^2\right>$,  $\left<
\overrightarrow{u}^2\right>$, or both vanish.

\section{Information entropy balance in  Smoluchowski diffusion process}

Remembering that in  the standard Brownian motion,
essentially the same mathematical formalism  applies  to a single
particle and to a statistical ensemble of identical noninteracting
Brownian particles, we shall adopt to our
purposes basic tenets of  so-called thermodynamic formalism of
isothermal diffusion processes, \cite{qian,qian1,qian2} (see also
 \cite{hatano} and \cite{rubi}), originally introduced in connection with  nonequilibrium
thermodynamics of single macromolecules immersed in an ambient
fluid at a constant temperature, and promoted in
\cite{qian} to the status of "stochastic macromolecular
mechanics".

Let us discuss in more detail  Eq.~(\ref{major}) for  the differential entropy
balance which is extremely  persuasive  in  the special case of Smoluchowski diffusions.
 Indeed, then
 $\overrightarrow{b} \doteq  \overrightarrow{F}/( m\beta )$ stands for an externally
 acting force, capable of performing a mechanical work  which in turn may  be converted into heat.
 We refer to the  standard phase-space conceptual background, \cite{huang,rubi}.

According to \cite{qian,qian1,qian2} (we adjust their  framework and notation to our purposes),
close to   equilibrium,   one expects the information   entropy  to decrease in the course of the
 Smoluchowski  diffusion process. The mean  rate of the entropy loss per unit of mass, equals:
\begin{equation}
{\frac{d{\cal{Q}}}{dt}} \doteq  {\frac{1}D} \int {\frac{1}{m\beta }}
\overrightarrow{F} \cdot \overrightarrow {j} d^3x =
{\frac{1}D} \langle \overrightarrow{b}\cdot
 \overrightarrow{v}
 \rangle \, .
 \end{equation}

That can be rewritten otherwise: $k_BT \dot{{\cal{Q}}}= \int \overrightarrow{F}\cdot
\overrightarrow{j} d^3x$, where $T$ is the temperature of the bath.
In the formal thermodynamical lore, we deal here with the time rate at which the mechanical work
is being dissipated  into thermal environment in the form of (removed)  heat.
 Let us point out that
this interpretation is surely true under equilibrium conditions \cite{qian}.  In general, far from
equilibrium, the sign of $d{\cal{Q}}/dt$  remains indefinite and  may refer to heat absorption (if negative)
instead of heat removal.

The nonnegative term in  Eq.~(\ref{major}) can be consistently  interpreted, c.f.  \cite{qian},
 as the measure of the  entropy  gain per unit of time  by the diffusion process. Accordingly, we have:
\begin{equation}
\frac{d{\cal{S}}}{dt} = \frac{d{\cal{S}}_{gain}}{dt}  - {\frac{d{\cal{Q}}}{dt}}  [\label{balanceeq}
\end{equation}
where  $d{\cal{S}}_{gain}/{dt} \doteq (1/D) \left<\overrightarrow{v}^2\right>$. If the entropy gain is
counterbalanced by heat removal, we may  have $d{\cal{S}}/dt = 0$.

Let us mention  that our  "entropy gain" is named  "entropy production" in Ref.~\cite{qian}.
In the earlier literature on the subject, \cite{ruelle}, the entropy production name  has been reserved
to  the accumulating  entropy  surplus which is  being  removed from the system under consideration to the
environment. In our discussion, just to  the contrary,  the information entropy appears to be pumped into
the system (e.g. the diffusion process) instead of being removed.

The relationship:
\begin{equation}
\overrightarrow{j} \doteq  \rho \, D \overrightarrow{F}_{th}
\end{equation}
defines a thermodynamic force $\overrightarrow{F}_{th} $  associated with the  Smoluchowski diffusion:
 \begin{equation}
 k_BT\, \overrightarrow{F}_{th}  = \overrightarrow{F} - k_BT\, \overrightarrow{\nabla } \ln \rho
 \doteq - \overrightarrow{\nabla } \Psi  \, .
  \end{equation}
Notice that
\begin{equation}
\overrightarrow{v} = - (1/m\beta ) \overrightarrow{\nabla  } \Psi  \, .
\end{equation}
In the absence of external force (free Brownian motion), we obviously get $D \overrightarrow{F}_{th} =
 -  \overrightarrow{u} $, $\dot{\cal{Q}}=0$  and $\dot{\cal{S}} = \dot{\cal{S}}_{gain}$, hence delocalization
 coincides with the "diffusion of probability".

The  mean value of the potential
\begin{equation}
\Psi = V + k_BT \ln \rho
\end{equation}
 of the thermodynamic force
defines the obvious diffusion process analogue of the   Helmholtz free energy:
\begin{equation}
\left< \Psi \right> = \left< V\right> - T\, {\cal{S}}_G \label{Helmholtz}
\end{equation}
where the dimensional version of  information entropy  ${\cal{S}}_G \doteq k_B {\cal{S}}$
has been introduced  (actually, it is a  direct analog of the Gibbs entropy).
The expectation value  of the mechanical force potential $ \left< V\right>$
plays here the role of the  mean internal energy.

By assuming that $\rho V \overrightarrow{v}$  vanishes at integration volume boundaries (or infinity), we
easily get the time rate of Helmholtz free energy:
\begin{equation}
{\frac{d}{dt}}  \left< {\Psi } \right> = - k_BT \dot{{\cal{Q}}}  -
T \dot{\cal{S}}_G
\end{equation}
where $k_BT \dot{{\cal{Q}}}= \int \overrightarrow{F}\cdot
\overrightarrow{j} d^3x$  and
$T \dot{\cal{S}}_G = \int (k_BT \overrightarrow{F}_{th} -  \overrightarrow{F}) \cdot \overrightarrow{j}d^3x$.
In view of Eq.~(\ref{balanceeq}) we  get
\begin{equation}
{\frac{d}{dt}} \left< {\Psi } \right>  = - (m\beta )
 \left<\overrightarrow{v}^2\right>
\end{equation}
which is  either  negative or vanishes.
 Therefore, the   Helmholtz free energy either remains constant in time
or decreases as a function of time.

 In the presence of external forces
this property quantifies a possible asymptotic approach towards a
minimum corresponding to an invariant density of the process.
Indeed, a  particular example of an equilibrium (invariant) density reads
 $\rho (x) = (1/Z) \exp(-V/k_BT) $,
where $Z=\int \exp(-V/k_BT)\, dx $. Such $\rho $  sets the  pertinent  minimum  of $\langle \Psi \rangle $
at   $ \langle \Psi \rangle  =  \Psi = -k_BT \ln  Z$.
This corresponds to   $\Psi = V + k_BT \ln \rho = const $ and thus trivially implies
  $\overrightarrow{\nabla } \Psi =\overrightarrow{0} = \overrightarrow{v}$.

One should be aware that  an invariant density as
well may  not exist: in case of free Brownian motion there is no
invariant density.

\section{Localization toolbox:  Shannon entropy and Fisher information}

For simplicity all of our further discussion will be restricted  to one
space dimension.

Let us  consider  the Gaussian
probability density  on the real line $R$ as a reference  density function:
 $\rho (x)= [1/(\sigma \sqrt{2\pi }] \exp[- (x-x_0)^2/2\sigma ^2]$.
Among \it all \rm  one-dimensional distribution
functions  $\rho (x)$   with a finite   mean,  subject
to the  constraint  that the standard deviation is fixed at $\sigma $, it is
the Gauss function  with half-width $\sigma $ which sets  a maximum  of the
 differential entropy, \cite{shannon}.  For the record, let us add that if
 only the mean is  given for  probability density functions on $R$, then there is
  no maximum entropy distribution in their set.

 The differential entropy of the Gauss density  has  a simple analytic form, independent of the mean
value $x_0$ and  maximizes an inequality:
\begin{equation}
 {\cal{S}}(\rho ) \leq  {\frac{1}{2}} \ln \,(2\pi e \sigma ^2) \, . \label{bound}
\end{equation}
This imposes a useful bound upon the so-called   entropy power, \cite{shannon}:
\begin{equation}
{\frac{1}{\sqrt{2\pi e}}} \exp [{\cal{S}}(\rho )] \leq  \sigma  \label{power}
\end{equation}
with an obvious bearing on the spatial localization  of the  density $\rho $,
 hence spatial (un)certainty of position measurements.   We can say that almost surely,  with
 probability  $0.998$, the probability is concentrated within the interval of the
length $6\sigma $ which is centered about the mean value $x_0$  of the Gaussian density
 $\rho $.

 The Shannon  entropy of an arbitrary  continuous  probability density  is unbounded form below
  and from above, but  in the  subset of  all
  densities with a finite mean and  a fixed variance $\sigma ^2$, we  actually have an upper bound
  set  by Eq.~(\ref{bound}).
Note that  not only for   small, but also  for relatively large   mean deviation
values  $\sigma < 1/\sqrt{2\pi e} \simeq  0.26 $ the  differential entropy  ${\cal{S}}(\rho )$  becomes
  negative.

Let us discuss to what extent, the Shannon  entropy can be viewed as
 a measure of   localization in the  configuration space
 of the   dynamical system.

Let us consider a one-parameter family of probability densities $\rho _{\alpha }(x)$
  on $R$ whose first (mean) and second  moments (effectively, the variance) are finite.
The parameter-dependence is here not completely arbitrary and we assume standard
regularity properties that allow to differentiate various
functions  of $\rho _{\alpha }$ with respect to  the parameter
 $\alpha $ under the sign of an  (improper) integral.

Namely, let us denote  $ \int x\rho _{\alpha }(x) dx = f(\alpha )$
and $\int x^2\rho _{\alpha } dx < \infty $. We demand that as a function of
$x\in R$, the  modulus of the partial derivative
$\partial \rho _{\alpha }/ \partial \alpha $ is  bounded by a function $G(x)$ which
together with  $x G(x)$ is integrable on $R$.  This implies,
the existence of $\partial f /\partial \alpha $ and an important inequality:
\begin{equation}
\int (x- \alpha )^2 \rho _{\alpha } dx \cdot \int \left(
 {\frac{\partial ln \rho _{\alpha }}{\partial \alpha }}
 \right)^2 \rho _{\alpha } dx  \label{cramer}
 \geq
 \left( \frac{df(\alpha )}{d\alpha }\right)^2
\end{equation}
directly resulting from
\begin{equation}
{\frac{d f}{d\alpha }} =
\int [(x - \alpha ) \rho _{\alpha }^{1/2}] [{\frac{\partial (\ln \rho _{\alpha })}
{\partial \alpha }} \rho _{\alpha }^{1/2}] dx
\end{equation}
via the standard Schwarz inequality, \cite{cramer}.
The  equality  appears  if $\rho _{\alpha }(x)$ is the  Gauss function with
 mean value $\alpha $.

At this point let assume that the mean value of $\rho _{\alpha }$ actually
  equals $\alpha $ and we fix at  $\sigma ^2$ the  value
   $\langle (x-\alpha )^2\rangle = \langle x^2 \rangle -  \alpha ^2 $
of  the variance  (in fact, standard deviation
from the mean value) of the probability density $\rho _{\alpha }$.
The previous  inequality   now  takes the  familiar  form:
\begin{equation}
 {\cal{F}}_{\alpha } \doteq  \int   {\frac{1}{\rho _{\alpha }}} \left({\frac{\partial
 \rho _{\alpha}}{\partial \alpha }}
\right)^2\, dx
\geq {\frac{1}{\sigma ^2}} \label{Fisher}
\end{equation}
where  an integral
on the left-hand-side is the so-called Fisher information of $\rho _{\alpha }$,
 known to   appear  in various problems of   statistical estimation
 theory, as well as an ingredient of a number of   information-theoretic inequalities.
  In view  of  ${\cal{F}}_{\alpha } \geq 1/\sigma ^2$,
 we realize that the Fisher  information  is   more sensitive indicator of the probability
 density  localization   than the entropy power, Eq.~(\ref{power}).

Let us  define $\rho _{\alpha }(x) \doteq \rho (x-\alpha )$. Then,
the Fisher information  can be readily  transformed  to the conspicuously
quantum mechanical form  (up to a factor  $D^2$ with $D=\hbar /2m$):
\begin{equation}
{\frac{1}{2}} {\cal{F}}_{\alpha } =   {\frac{1}{2}} \int   {\frac{1}{\rho }} \left({\frac{\partial \rho }{\partial x }}
\right)^2\, dx  = \int \rho \cdot  {\frac{u^2}{2}} dx = - \langle Q \rangle  \label{Fisher1}
  \end{equation}
 where  $u\doteq  \nabla \ln \rho $ (up to a factor $D$) represents   an osmotic velocity field,
 \cite{nelson,gar}, and an average $\langle Q\rangle =\int  \rho \cdot Q dx$ is carried out with
 respect to the function
 \begin{equation}
Q= 2 {\frac{\Delta \rho ^{1/2}}{\rho ^{1/2}}}  \, . \label{potential}
\end{equation}
As a consequence of Eq.~(\ref{Fisher}), we have $- \langle Q\rangle \geq 1/2\sigma ^2$ for all
relevant probability densities with  variance $\sigma ^2$.

An important inequality, valid under an assumption  $\rho _{\alpha }(x) = \rho (x- \alpha )$,
has been proved in \cite{stam}:
\begin{equation}
{\frac{1}{\sigma ^2}} \leq    (2\pi e) \, \exp[- 2 {\cal{S}}(\rho )] \leq   {\cal{F}}_{\alpha }
\end{equation}
It tells  us that the lower bound for the Fisher information is in fact  given a  sharper  form  by means
of the (squared) inverse entropy power. Our  two information measures  appear to be correlated.

Under an additional decomposition/factorization  ansatz (of the  quantum mechanical  $L^2(R^n)$  provenance) that
$\rho (x) \doteq |\psi |^2(x)$,     where a real or complex function  $\psi  =\sqrt{\rho } \exp(i\phi )$
is a normalized  element of
$L^2(R)$, another important inequality holds true,  \cite{stam}:
\begin{equation}
{\cal{F}}_{\alpha }  =  4 \int \left({\frac{\partial \sqrt{\rho }}{\partial x}} \right)^2 dx
\leq   16 \pi ^2 \tilde{\sigma }^2 \, ,
\end{equation}
provided  the Fisher information takes finite values.
Here, $\tilde{\sigma }^2$ is  the variance of
the "quantum mechanical momentum canonically conjugate
to the position observable", up to (skipped) dimensional factors.
In the above, we have exploited the
Fourier transform $\tilde{\psi } \doteq ({\cal{F}} \psi)$  of $\psi $ to arrive at
$\tilde{\rho }\doteq |\tilde{\psi }|^2$   whose variance the above
$\tilde{\sigma }^2$ actually is.

 Let us point out that the Fisher information ${\cal{F}}(\rho )$ may blow up to infinity
under a number of circumstances: when $\rho $ approaches the Dirac delta behavior,
if $\rho $    vanishes over some  interval in $R$  or   is discontinuous.
We observe that  ${\cal{F}}>0$ because  it  may   vanish
only when  $\rho $ is constant  everywhere on $R$, hence  when $\rho $ is \it not  \rm
 a probability density on $R$.

In view of  two previous inequalities,  we find out that not only the Fisher
information, but also  an entropy power  may be  bounded from below and  above. Namely, we have:
\begin{equation}
{\frac{1}{\sigma ^2}} \leq {\cal{F}}_{\alpha } \leq  16\pi ^2 \tilde{\sigma }^2 \label{in1}
\end{equation}
which implies $1/2\sigma ^2 \leq  -\langle Q  \rangle \leq 8\pi ^2 \tilde{\sigma }^2 $  and furthermore
\begin{equation}
{\frac{1}{4\pi \tilde{\sigma }}} \leq {\frac{1}{\sqrt{2\pi e}}}\, \exp[{\cal{S}}(\rho)]  \leq  \sigma \, .
\label{in2}
\end{equation}
  Most important outcome of Eq.~({\ref{in2}) is that the differential
 entropy ${\cal{S}}(\rho) $  typically  may be expected to be a well behaved   quantity:
  with finite  lower and upper bounds. A standard statement in this regard is:  Shannon entropy of
  a continuous probability density is neither bounded form below nor from above, \cite{shannon,sobczyk}.

\section{Dynamics of uncertainty: mean energy  versus  localization}

When multiplied by $D^2$, a   potential-type function
 $Q = Q(x,t)$,  c.f. Eq.~(\ref{potential}) notoriously appears in the
hydrodynamical formalism    of quantum mechanics  as the  so-called  de Broglie-Bohm quantum potential
($D=\hbar /2m$), \cite{gar,czopnik}. It appears as well  in   the corresponding formalism
for diffusion-type  processes, including the standard Brownian motion
(then,  $D=k_BT/m\beta $, see e.g.   \cite{geilikman,gar,gar1,skorobogatov,gar2}.
We have:
\begin{equation}
Q=2D^{2}{\frac{\Delta \rho ^{1/2}} {\rho ^{1/2}}} =
{\frac{1}2} u^2 + D \nabla \cdot u \,  \label{potential1}
\end{equation}
and it is instructive to notice that  the   gradient of $Q$   trivially  appears
(i.e. merely as a consequence of the heat equation, \cite{geilikman,gar,czopnik})
  in the   hydrodynamical (momentum) conservation law  appropriate for the free Brownian motion:
\begin{equation}
\partial _t {v} + (v \cdot
{\nabla }) {v} = - {\nabla }Q \, . \label{free}
\end{equation}

We assume, modulo restrictions upon drift functions \cite{eberle,qian1}, that
the Smoluchowski dynamics  can be resolved  in terms of (possibly non-unique) Markovian diffusion-type processes.
Then, the following  compatibility equations follow in  the form of  local (hydrodynamical)   conservation
laws for the diffusion process, \cite{gar,czopnik}:
\begin{eqnarray}
\partial _{t}\rho + {\nabla }(
{v} \rho )%
 &=&0 \\
(\partial _{t} + {v}
\cdot {\nabla }) {v}%
 &=& {\nabla } ( \Omega -Q)\,  \label{law}
\end{eqnarray}
where, not to confuse this notion with the previous force field potential  $V$,  we denote by $\Omega  (x)$
  the so-called volume potential for the process:
\begin{equation}
\Omega =\frac{1}{2}\left( \frac{F}{\ m\beta }\right) ^{2}+D{\nabla
}\cdot \left( \frac{F}{\ m\beta }\right) \, .  \label{Omega}
\end{equation}
Obviously the free Brownian  law,  Eq. (\ref{free}),  comes out as the  special case.

In the above (we use  a short-hand notation
$v \doteq v(x,t)$):
\begin{equation}
v \doteq   b - u
 = \frac{%
 F}{\ m\beta }- D\frac{\nabla
\rho }{\rho }
\end{equation}
 defines  the  current velocity of Brownian particles in external force field.
 This formula allows us  to   transform the continuity equation into the
 Fokker-Planck equation  and back.

By considering   $( - \rho )(x,t)$ and
 $s(x,t)$, such that
$v= {\nabla }s$, as canonically conjugate
 fields,  we can  invoke the variational calculus, \cite{skorobogatov,reginatto}.
 Namely, one may
   derive  the continuity (and thus Fokker-Planck) equation together
 with  the Hamilton-Jacobi type equation (whose gradient implies the hydrodynamical
 conservation law Eq.~(\ref{law})):
\begin{equation}
\partial _ts +\frac{1}2 ({\nabla }s)^2 - (\Omega -  Q) = 0
\label{jacobi}\, ,
\end{equation}
 by means of the extremal (least, with fixed end-point variations) action
principle
 involving  the (mean) Lagrangian:
\begin{equation}
{\cal{L}} = - \int \rho\left[ \partial _t s   +  {\frac{1}2}
({\nabla }s)^2  -
\left({\frac{{u}^2}2} + \Omega \right) \right]
dx\, .
\end{equation}

The related  Hamiltonian  (which is  the mean energy of the diffusion process  per unit of mass) reads
\begin{equation}
{\cal{H}} \doteq \int  \rho \cdot \left[ {\frac{1}2}({\nabla }
s)^2  -
\left( {\frac{{u}^2}2} + \Omega \right ) \right] \, dx \label{energy}
\end{equation}
i. e.
$${\cal{H}} = (1/2) (\left< {v}^2\right> - \left< {u}^2\right>)  -
\left<\Omega \right> \, .
$$

We can evaluate an expectation value of Eq. (\ref{jacobi}) which  implies an
identity ${\cal{H}} = - \left< \partial _ts \right>$.  By invoking the Smoluchowski diffusion and
thus  Eq.~(\ref{Helmholtz}),  with the  time-independent $V$,  we  arrive at
\begin{equation}
 \dot{\Psi } = {\frac{ k_BT}{\rho }}  \nabla (v\rho )
\end{equation}
  whose   expectation value  $\langle \dot{\Psi } \rangle$, in view of
   $v\rho =0$  at   the integration volume boundaries,  identically  vanishes.
   Since   $v= - (1/m\beta ) \nabla \Psi $,  we  define
   \begin{equation}
   s(x,t)\doteq (1/m\beta ) \Psi (x,t)  \Longrightarrow  \left< \partial _ts \right>=0
   \end{equation}
so that ${\cal{H}} \equiv 0$ identically.

 We have  thus  arrived at the  following     interplay between the mean energy, localization
 and the information entropy  gain:
  \begin{equation}
{\frac{D}2} \left( {\frac{dS}{dt}}\right)_{gain} = \int  \rho \, \left(
{{\overrightarrow{v}}^2\over 2} \right) \, dx =
  \int  \rho \, \left(
{{\overrightarrow{u}}^2\over 2} + \Omega \right) \, dx  \geq 0  \, ,
\end{equation}
generally valid for Smoluchowski processes  with non-vanishing diffusion currents.

By recalling the notion of the Fisher information
Eq.~(\ref{Fisher1}) and setting ${\cal{F}} \doteq  D^2 {\cal{F}}_{\alpha }$,  we can rewrite
the above formula as follows:
\begin{equation}
{\cal{F}}  =  \langle v^2 \rangle - 2 \langle
\Omega \rangle \geq  0   \label{interplay}
\end{equation}
 where  $
{\cal{F}}/2 = - \langle Q \rangle >0$  holds true for probability
densities with  finite mean and variance.

We may evaluate directly the localization/uncertainty dynamics of the Smoluchowski process,
by recalling that the Fisher information ${\cal{F}}/2$ is the localization measure,
which for probability densities with   finite mean value and
 variance $\sigma ^2$ is  bounded from below by $1/\sigma ^2$.

Namely, by exploiting the  hydrodynamical  conservation laws  Eq.~(\ref{law})  for
the Smoluchowski process   we get:
\begin{equation}
\partial _t (\rho {v}^2) = - {\nabla}\cdot
 [(\rho  {v}^3)] - 2\rho {v}\cdot
 {\nabla }(Q - \Omega) \, .
 \end{equation}
We assume to have  secured conditions allowing to take a derivative under an
indefinite integral, and  assume  that  of $\rho  {v}^3$  vanishes
 at the integration volume  boundaries. This implies
 the following expression for the time derivative of $\left< {v}^2\right>$:
\begin{equation}
 {\frac{d}{dt}}{\left< {v}^2\right>} = 2 \left< {v}\cdot {\nabla }
(\Omega - Q) \right> \, .
\end{equation}

Proceeding in the same vein, in view of $\dot{\Omega } =0$,  we find that
\begin{equation}
{\frac{d}{dt}} \langle \Omega \rangle =  \langle v \cdot \nabla \Omega \rangle
\end{equation}
and so the equation of motion for ${\cal{F}}$ follows:
\begin{equation}
{\frac{d}{dt}} {\cal{F}} = {\frac{d}{dt}} [ \langle v^2\rangle -
2\langle \Omega \rangle  ]=
   -  2 \langle v\cdot \nabla Q \rangle \, .
\label{Fishdynamics}
\end{equation}

Since  we have $\nabla Q = \nabla P/\rho $ where $P=D^2\rho\, \Delta \ln \rho $,
the previous equation takes the form $\dot{\cal{F}} =  - \int \rho v \nabla Q dx
=  - \int v \nabla P dx$,
which is an analog of the familiar expression for the
power release ($dE/dt = F\cdot v$, with $F=-\nabla V$) in classical mechanics.

This  should  be compared with our  previous discussion of the "heat dissipation"
 term. Indeed, $\dot{\cal{F}} = \int j\cdot (-2\nabla Q)dx$, while  the expression
 for the heat dissipation rate had the form
 $k_BT\dot{\cal{Q}}= \int j \cdot (-\nabla V)dx$.

Let us notice that $\dot{\cal{F}}>0$  would tell us  that the localization improves,
 clearly at the expense of  the energy supply  (power injection) from the environment.
 $\dot{\cal{F}}< 0$ indicates a localization decay and  corresponds to the energy
 absorption (power release) by the environment.

We may typically  expect the  decrease of  the the localization measure ${\cal{F}}$ and the continual
 energy/heat  absorption by the Smoluchowski  diffusion process.
This effect can be attributed to the active role of the   thermal environment which  generally
  leads to a delocalization of the initially localized probability density, unless
  the invariant measures enter the game.
 The power release   complies with the identity ${\cal{H}}\equiv 0$ since "obviously"
  the diffusion process proceeds in an open system.  The latter  property  should be
  contrasted with the behavior  of so-called  finite energy diffusions, \cite{nelson,gar,gar3}.\\

{\bf  Acknowledgement:}  Dedicated to Professor Andrzej Fuli\'{n}ski, with admiration.\\
The paper has been supported by the Polish Ministry of Scientific Research grant
No PBZ-MIN-008/P03/2003.

\end{document}